# Title: Detection of Carbon Monoxide and Water Absorption Lines in an Exoplanet Atmosphere

Authors: Quinn M. Konopacky[1,3*†], Travis S. Barman[2], Bruce A. Macintosh[3], Christian Marois[4]

Affiliations:

[1]Dunlap Institute for Astronomy and Astrophysics, University of Toronto, 50 St. George Street, Toronto, ON M5S 3H4, Canada.

[2]Lowell Observatory, 1400 West Mars Hill Road, Flagstaff, AZ 86001, USA.

[3]Lawrence Livermore National Laboratory, 7000 East Avenue, Livermore, CA 94550, USA.

[4]National Research Council of Canada, Dominion Astrophysical Observatory, 5071 West Saanich Road, Victoria, BC V9E 2E7, Canada.

*Correspondence to: konopacky@di.utoronto.ca.

†Dunlap Fellow.

**Abstract**: Determining the atmospheric structure and chemical composition of an exoplanet remains a formidable goal. Fortunately, advancements in the study of exoplanets and their atmospheres have come in the form of direct imaging – spatially resolving the planet from its parent star - which enables high-resolution spectroscopy of self-luminous planets in Jovian-like orbits. Here, we present a spectrum with numerous, well-resolved, molecular lines from both water and carbon monoxide from a massive planet orbiting less than 40 AU from the star HR 8799. These data reveal the planet's chemical composition, atmospheric structure, and surface gravity confirming that it is indeed a young planet. The spectral lines suggest an atmospheric carbon-to-oxygen ratio greater than the host star's, providing hints about the planet's formation.

HR 8799 is a young (~30 Myr, *1*) early type (A5 to F0) star ~130 light years from the Sun. Near-infrared direct imaging using adaptive optics revealed four planets around it (*2, 3*). A number of studies have broadly characterized these planets, showing that all four have masses between 5 and 10 times that of Jupiter, effective temperatures between 900 and 1200K (still hot from the gravitational energy released during their formation) and near-infrared colors that are redder than initially expected for their mass and temperature ranges. The red colors are best explained by the presence of iron and silicate atmospheric clouds. These clouds are normally located below the photosphere in old field brown dwarfs with temperatures less than 1400 K, but are believed to persist at cooler temperatures in young planets where surface gravities are much lower (*4-6*). Low surface gravity may also contribute to extreme deviations from equilibrium concentrations of carbon monoxide (CO) and methane ($CH_4$), potentially explaining the lack of strong $CH_4$ absorption (*7*) previously anticipated in young planets. The formation mechanism of the HR 8799 planets – and indeed all exoplanets – remains uncertain, with both global disk instabilities (*8*) and bottom-up core accretion (*9, 10*) being proposed. One possible way to differentiate between these is through planetary composition. Planets that form through disk instabilities will track the bulk abundance of the original star-forming material, whereas core-accretion planets may be enhanced or depleted in some elements.

Further inferences on the chemical make up and surface gravity of the HR 8799 planets, and exoplanets in general, have thus far been limited by poor spectral resolution (typically $\lambda/\Delta\lambda$

~10-100). Here we present a moderate resolution ($\lambda/\Delta\lambda$ ~4000) spectrum of the planet HR 8799c.

Observations of HR 8799c were obtained in 2010 and 2011 with the Keck II 10-m telescope, using the integral field spectrograph, OSIRIS (*11*) in conjunction with the facility adaptive optics system (*12*). These observations were obtained in the K band (1.965 μm to 2.381 μm) at the finest spatial sampling provided by OSIRIS (0.02" per spaxel). A total of 5.5 hours of observations were obtained, consisting of 33 exposures, each 600 seconds long. Raw data were processed through sky and dark subtraction, bad pixel removal, telluric calibration from an A0 stellar standard, and transformed into datacubes (*13*). Each has ~1600 spectral channels with a 0.32" by 1.28" field of view at a nominal spectral resolution ($\lambda/\Delta\lambda$) ~4000 per spatial location.

Scattered starlight in the form of an interference pattern of speckles makes identification of the planet challenging. We used speckle modeling and subtraction (*14*) for spectral extraction. A key insight is that the residual speckle noise has a characteristic spectral signature – generally broad (500 Å) features passing through the planetary spectrum. This produces highly correlated noise in a low-resolution spectrum over the full band pass, but does not affect the detectability of individual narrow spectral lines. Spectra were extracted from each of the 33 datacubes and median combined into a single spectrum with an average signal to noise ratio (SNR) per channel of ~30 (*15*).

Only a handful of objects with mass and age similar to HR 8799c are known and broad K-band spectra are available for only a few of these, all with resolutions substantially lower than 4000. We compared the extracted spectrum of HR 8799c to the spectrum of a ~5 $M_{Jup}$ brown dwarf, 2MASS J12073346-3932539b (2M1207b, *16*) and the spectrum of HR 8799b (*14*), all three binned to the same resolution (Fig. 1). HR 8799c falls between these two in appearance and, like 2M1207b, has an obvious CO bandhead at ~2.29 μm. The spectral morphology and colors of HR 8799b and 2M1207b also imply the presence of dust clouds in their atmospheres (*14, 17-18*). The rough morphological similarity between HR 8799c and these objects implies that it, too, has a cloudy atmosphere. The inference of atmospheric clouds is consistent with results from studies based on the photometry of HR 8799c (*2, 17-21*), which find that thick dust clouds are necessary to explain its spectral energy distribution.

With the speckles mostly removed, the long-exposure (5.5 hours) data are sufficient to examine the spectrum at the nominal resolution of OSIRIS (Fig. 2). Numerous spectral lines are immediately obvious, particularly those corresponding to strong CO (2,0) and (3,1) band heads at 2.29 and 2.32 μm. The remaining absorption lines are expected to be due to $H_2O$.

To further verify that the features are in fact molecular lines and not residual noise, the spectrum was low-pass filtered. Filtering effectively removes any residual speckle artifacts in the spectrum as well as the planet's continuum, leaving behind only the absorption lines (Fig. 3; *15*). Because the scattered light has roughly the spectrum of HR 8799A -- a nearly featureless A-type star – modulated at 500 Å scales by the speckle process, all remaining narrow spectral lines are from the planet. Although many CO and $H_2O$ lines are visible, no $CH_4$ lines are identified. We cross-correlated the low-pass filtered planetary spectrum with spectral templates of pure $H_2O$, CO and $CH_4$. Given the incompleteness of methane molecular line data, we used three popular, yet fairly distinct, $CH_4$ lists as templates (*22-25*) (Fig. 3). The cross-correlation confirms that all lines detected in the low-pass filtered observations are attributable to CO and $H_2O$.

Two quantities that greatly influence a planet's atmosphere are effective temperature and gravity. The effective temperature of HR8799c is likely between 900 and 1300K based on its observed bolometric luminosity ($-4.8 < \log(L/L_\odot) < -4.6$) (*2*) and the range of radii predicted by formation and evolution models (1 to 1.5 Jupiter-radii). Dynamical stability requirements of the planetary system require all four planets to be less than 10 $M_{Jup}$ (*2, 26*). This mass limit, combined with the above radius range, implies a surface gravity below 250 m s$^2$. Furthermore, the planetary system's youth (~30 Myrs, *1, 27*) and cooling tracks for giant planets indicate a surface gravity between ~30 m s$^{-2}$ and ~250 m s$^{-2}$.

The upper limit we set on rotation ($v\sin i$) for HR 8799c is ~40 km s$^{-1}$, roughly the resolution of our data, indicating that the cumulative effect of broadening by gravity and rotation is modest. We compared the grid of model spectra from (*14*), convolved to the nominal OSIRIS resolution (~5 Å) and low-pass filtered in the same manner as the observed spectrum, to the drop in flux across the CO (2,0) bandhead, a feature known to be gravity sensitive (*15*). Surface gravities of ~100 m s$^{-2}$ or less, consistent with the range based on the system's youth and dynamical stability, provide the best match.

We further refined the effective temperature using the K-band spectrum and the surface gravity limit (< 100 m s$^{-2}$). The K-band spectrum has uniformly characterized systematics and uncertainties, and contains key features - the overall shape and the CO bandhead depth. Fitting the binned, low-resolution, spectrum with a subset of the model grid having $\log(g) = 3.5 - 4.0$ (cgs units) results in $T_{eff} = 1100 \pm 100$ K (*15*). The best matching model (Fig. 1 and 2) produces a cross-correlation peak that exceeds the correlations described above (Fig. 3). The model that best matches our K-band spectrum also compares well to the available photometry, including recent photometry (*18*), with $\chi^2$ values equal to or smaller than previously published fits (*2, 17-21*). Previous results have converged on this temperature and gravity range by fitting photometric observations with broad wavelength coverage (*17*). For the remainder of the analysis, $T_{eff} = 1100 \pm 100$K and $\log(g) = 3.5 - 4.0$ (cgs units) are adopted as the most plausible values. The implied mass and age ranges for these values are ~3-7 $M_{Jup}$ and ~3-30 Myr (hot start models, *28*).

Our data require the mixing ratio of $CH_4$ to be less than ~10$^{-5}$ (assuming the methane templates are accurate for the strongest lines, *25*). At $T_{eff} \sim 1100$K, such a low $CH_4$ mixing ratio indicates that vertical mixing has quenched the CO and $CH_4$ mole-fractions at depths corresponding to their maximum and minimum values, respectively. Though many details concerning vertical mixing of CO and its impact on chemical equilibrium are poorly understood, it is well understood that reestablishing equilibrium CO/$CH_4$ after CO has been mixed into a normally $CH_4$-rich photosphere can take 10$^5$ to 10$^9$ years (*29*). Therefore, even modest vertical mixing will quench the CO and $CH_4$ (and $H_2O$) in the atmosphere where the chemical timescales exceed the mixing timescales (*30*). Deep quenching requires short mixing timescales ($t_{mix} \sim Hp^2/Kzz$, where Hp is the pressure scale height and Kzz is the eddy diffusion coefficient), justifying the use of a large Kzz = 10$^8$ cm$^2$ s$^{-1}$ in our models.

Of course, the final quenched mixing ratios of CO, $CH_4$ and $H_2O$ all depend on the bulk atmospheric elemental abundances that, so far, have been assumed to be equal to that of the Sun (*31-32*). The planetary atmospheric abundances may differ from those of the host star depending on how the planet formed, adding additional significance to chemical composition.

Under gravitational instability (GI), planets are the product of disk instabilities that gravitationally collapse and, while there are several important stages of the collapse, the planet interior and atmosphere are formed simultaneously with a stellar composition (ultimately that of the molecular cloud from which the system formed). Deviations from a stellar composition are possible through post-formation capture of solid material; however, it has been shown that if the HR 8799 planets are indeed massive (> 3 $M_{Jup}$) the timescale for substantial planetesimal capture is too short (especially at their current separations) to allow for compositions very different from stellar (*33*). If the HR8799 planets formed by GI, the spectrum of HR8799c should indicate a stellar composition and in particular a stellar C/O (assumed equal to that of the Sun for HR8799, (*34*).

Under core accretion (CA), planets form in a multi-step process involving the initial formation of a core (on the order of 10 Earth-masses of heavy elements) followed by runaway accretion primarily of gas supplied by the disk. When the disk is no longer able to supply a substantial amount of material the newly formed planet is isolated from what remains of the disk. The final atmospheric composition of a planet formed by CA depends on its location within the disk and the contribution of solids during the runaway accretion phase. A variety of compositions are possible (including stellar C/O with sufficient solid accretion), but a non-stellar composition is highly likely for massive giant planets.

Within the disk that formed the HR8799 planets, there are three important boundaries: the $H_2O$ (~10 AU), $CO_2$ (~90 AU) and CO (~600 AU) frost lines. The planets currently orbit their star between 15 and 70 AU and, therefore, are all located between the $H_2O$ and $CO_2$ lines where the gas phase C and O abundances in the disk would have been reduced through the formation of ice and carbon and silicate grains (with CO and $CO_2$ remaining in the gas phase). Therefore, in the CA scenario, planetary atmospheres acquired through gas-only accretion will have sub-stellar C and O abundances, but super-stellar C/O because water ice is more abundant than carbon-bearing grains. A simple model of ice formation suggests the disk gas-phase C/O ~ 0.9 (*35*). Increasing the fraction of the atmosphere acquired by solid accretion can lead to super-stellar values of both C and O, with the C abundance rising more slowly than O, and an overall decrease in C/O (*35*). Between the $CO_2$ and CO ice-lines, the abundances follow a similar pattern.

To explore the consequences of non-stellar C and O abundances, we made a grid of planetary atmosphere models following (*14*), but using the C and O values predicted by (*35*) for different ratios of solid to gas accretion, with C/O ranging from 0.45 to 1 (Tables S1 and S2). Again, we assumed that the C and O abundances of HR 8799A are solar (*15, 32, 34*). We found the best fit from this grid of model atmosphere spectra by minimizing $\chi^2$ (Fig. 4). While a comprehensive range of C and O values, independent of any disk-chemistry model, have not been explored, the results from these fits suggest that C/O is certainly less than one and not substellar, but likely larger than solar/stellar with substellar C and substellar O.

While it is fairly straightforward to rule out the extrema of C/O ratios, understanding the uncertainty in our C/O measurement is less so. In addition to noise in the data, the uncertainties discussed above in temperature and surface gravity also contribute to measurement errors in C and O, as line depths are sensitive to both of these bulk parameters. To marginalize over the temperature and gravity uncertainties, we expanded the grid of atmosphere models with abundances given in Table S1 to include temperatures of 1000-1200 K and log(g) of 3.5-4.0. We then performed a Monte Carlo simulation by resampling the spectral data from Gaussian distributions with widths determined by the (uncorrelated) uncertainties for all wavelength channels. We fit each resampled spectrum using the model grid, and recorded the best-fit abundances. The resulting estimate (and uncertainty) for C/O is $0.65^{+0.10}_{-0.05}$, marginally greater than the assumed stellar ratio (~0.55).

Measuring abundances is complicated and model atmospheres have not been thoroughly tested for systematic under or overestimation of C/O in substellar objects. However given the dominance of $H_2O$, CO and $CH_4$ opacities in their atmospheres, large variations in C/O should be easier to discern in brown dwarfs and giant planets than in stars (*36*). Planet migration and chemical evolution within the disk can muddy conclusions based on composition (*37*), as can core dredging, which may be important in planets more massive than Jupiter (*38*). With these caveats in mind, the above analysis rules out a planetary atmosphere for HR8799c that formed by gas-only accretion during a CA process at its current location (C/O > 0.9) and marginally excludes an atmosphere that formed from extreme amounts of solid accretion (C/O < 0.6). Between these extreme predictions the picture is more complicated, but the enhanced C/O ratio and the depleted C and O levels tend to favor a history in which the planet formed via CA. In this case, after an initial solid core formed, the planet atmosphere accreted from material that was partially but not completely depleted of solid planetesimals. However, given the uncertainties in parameters like the stellar abundances (*15*), a GI formation scenario is not totally excluded for this system. Our work shows the power of high-resolution spectra, which allows molecular species to be seen directly through their individual absorption lines rather than inferred from overall spectral shapes that are more sensitive to model parameters.

**References and Notes:**

1. B. Zuckerman, J. H. Rhee, I. Song, M. S. Bessell, The Tucana/Horologium, Columba, AB Doradus, and Argus Associations: New members and dusty debris disks. *Astrophys. J.* **732**, 61-79 (2011).
2. C. Marois *et al.* Direct imaging of multiple planets orbiting the star HR 8799. *Science* **322**, 1348-1352 (2008).
3. C. Marois, B. Zuckerman, Q. M. Konopacky, B. Macintosh, T. Barman, Images of a fourth planet orbiting HR 8799. *Nature* **468**, 1080-1083 (2010).
4. S. A. Metchev, L. A. Hillenbrand, HD 203030B: An unusually cool young substellar companion near the L/T transition. *Astrophys. J.* **651**, 1166-1176 (2006).
5. D. Saumon, M. S. Marley, The evolution of L and T dwarfs in color-magnitude diagrams. *Astrophys. J.* **689**, 1327-1344 (2008).
6. D. C. Stephens *et al.* The 0.8-14.5 μm spectra of mid-L to mid-T dwarfs: Diagnostics of effective temperature, grain sedimentation, gas transport, and surface gravity. *Astrophys. J.* **702**, 154-170 (2009).
7. T. S. Barman, B. Macintosh, Q. M. Konopacky, C. Marois, The young planet-mass object 2M1207b: A cool, cloudy, and methane-poor atmosphere. *Astrophys. J.* **735**, L39-L43 (2011).
8. A.P. Boss, Giant planet formation by gravitational instability. *Science* **276**, 1836-1839 (1997).
9. J. B. Pollack *et al.* Formation of the giant planets by concurrent accretion of solids and gas. *Icarus* **124**, 62-85 (1996).
10. R.R. Rafikov, Constraint on the giant planet production by core accretion. *Astrophys. J.* **727**, 86-93 (2011).
11. J. Larkin *et al.* OSIRIS: A diffraction limited integral field spectrograph for Keck. *New Astron. Rev.* **50**, 362-364 (2006).
12. P. L. Wizinowich *et al.* Adaptive optics developments at Keck Observatory. *Proc. SPIE* **6272**, 627209-1-9 (2006).
13. A. Krabbe *et al.* Data reduction pipeline for OSIRIS, the new NIR diffraction-limited imaging field spectrograph for the Keck adaptive optics system. *Proc. SPIE* **5492**, 1403-1410 (2004).
14. T. S. Barman, B. Macintosh, Q. M. Konopacky, C. Marois, Clouds and chemistry in the atmosphere of extrasolar planet HR8799b. *Astrophys. J.* **733**, 65-82 (2011).
15. See the supplementary materials for more information.
16. J. Patience, R. R. King, R. J. de Rosa, C. Marois, The highest resolution near infrared spectrum of the imaged planetary mass companion 2M1207 b. *Astron. Astrophys.* **517**, 76-81 (2010).

17. M. S. Marley *et al.* Masses, radii, and cloud properties of the HR 8799 planets. *Astrophys. J.* **754**, 135-151 (2012).

18. A. J. Skemer *et al.* First light LBT AO images of HR 8799 bcde at 1.6 and 3.3 μm: New discrepancies between young planets and old brown dwarfs. *Astrophys. J.* **753**, 14-25 (2012).

19. T. Currie *et al.* A combined Subaru/VLT/MMT 1-5 μm study of planets orbiting HR 8799: Implications for atmospheric properties, masses, and formation. *Astrophys. J.* **729**, 128-147 (2011).

20. N. Madhusudhan, A. Burrows, T. Currie, Model atmospheres for massive gas giants with thick clouds: Application to the HR 8799 planets and predictions for future detections. *Astrophys. J.* **737**, 34-48 (2011).

21. R. Galicher, C. Marois, B. Macintosh, T. Barman, Q. Konopacky, M-band imaging of the HR 8799 planetary system using an innovative LOCI-based background subtraction technique. *Astrophys. J.* **739**, L41-L45 (2011).

22. L. S. Rothman, F. J. Martin-Torres, J-M. Flaud, Special issue on planetary atmospheres. *J. Quant. Spec. Rad. Trans.* **109**, 881 (2008).

23. C. Wenger, J. P. Champion, Spherical Top Data System (STDS) software for the simulation of spherical top spectra. *J. Quant. Spec. Rad. Trans.* **59**, 471-480 (1998).

24. R. Warmbier *et al.* Ab initio modeling of molecular IR spectra of astrophysical interest: application to $CH_4$. *Astron. Astrophys.* **495**, 655-661 (2009).

25. A. Borysow, C. Wenger, J.P. Champion, U.G. Jorgensen, Towards simulation of high temperature methane spectra. *Mol. Phys.* **100**, 3585-3594, (2003).

26. J. J. Sudol, N. Haghighipour, High-mass, four-planet configurations for HR 8799: Constraining the orbital inclination and age of the system. *Astrophys. J.* **755**, 38-46 (2012).

27. E. K. Baines *et al.* The CHARA array angular diameter of HR 8799 favors planetary masses for its imaged companions. *Astrophys. J.* **761**, 57-71 (2012).

28. I. Baraffe, G. Chabrier, T.S. Barman, F. Allard, P.H. Hauschildt, Evolutionary models for cool brown dwarfs and extrasolar giant planets. The case of HD 209458. *Astron. Astrophys.* **402**, 701-712 (2003).

29. B. Fegley, Jr., K. Lodders, Atmospheric chemistry of the brown dwarf Gliese 229B: Thermochemical equilibrium predictions. *Astrophys. J.* **472**, L37-L39 (1996).

30. M.D. Smith, Estimation of a length scale to use with the quench level approximation for obtaining chemical abundances. *Icarus* **132**, 176-184 (1998).

31. M. Asplund, N. Grevesse, A.J. Sauval, in Cosmic Abundances as Records of Stellar Evolution and Nucleosynthesis, Thomas G. Barnes III, Frank N. Bash, Eds. (Astronomical Society of the Pacific), p. 25. (2005).

32. The spectrum of HR 8799 – a classic Lambda Bootes star - shows significant depletion of some elements. However, its peculiar abundances have been attributed to its extremely shallow photosphere and reaccretion of gas depleted in refractory elements -- possibly associated with planet formation (*39*) – so we assume here that the planets formed from solar-abundance material.

33. R. Helled, R. Bodenheimer, Metallicity of the massive protoplanets around HR 8799 if formed by gravitational instability. *Icarus* **207**, 503-508 (2010).

34. K. Sadakane, λ Bootis-like abundances in the Vega-like, γ Doradus type-pulsator HD 218396. *Pub. Astron. Soc. Japan* **58**, 1023-1032 (2006).

35. K. I. Öberg, R. Murray-Clay, E. A. Bergin, The effects of snowlines on C/O in planetary atmospheres. *Astrophys. J.* **743**, L16-L20 (2011).

36. J. J. Fortney, On the carbon-to-oxygen ratio measurement in nearby Sun-like stars: Implications for planet formation and the determination of stellar abundances. *Astrophys. J.* **747**, L27-L31 (2012).

37. K. Lodders, Jupiter formed with more tar than ice. *Astrophys. J.* **611**, 587-597 (2004).

38. H. F. Wilson, B. Militzer, Rocky core solubility in Jupiter and giant exoplanets. *Phys. Rev. Lett.* **108**, 111101-1-4 (2012).

39. R. O. Gray, C. J. Corbally, A spectroscopic search for λ Bootis and other peculiar A-Type Stars in intermediate-age open clusters. *Astron. J.* **124**, 989-1000 (2002).

40. B. P. Bowler, M. C. Liu, T. J. Dupuy, M. C. Cushing, Near-infrared spectroscopy of the extrasolar planet HR 8799 b. *Astrophys. J.* **723**, 850-868 (2010).

41. C. Marois, R. Doyon, R. Racine, D. Nadeau, Efficient speckle noise attenuation in faint companion imaging. *Pub. Astron. Soc. Pac.* **112**, 91-96 (2000).

42. W. B. Sparks, H. C. Ford, Imaging spectroscopy for extrasolar planet detection. *Astrophys. J.* **578**, 543-564 (2002).

43. D. Lafrenière, C. Marois, R. Doyon, D. Nadeau, É. Artigau, A new algorithm for point-spread function subtraction in high-contrast imaging: A demonstration with angular differential imaging. *Astrophys. J.* **660**, 770-780 (2007).

44. M. C. Cushing *et al.* Atmospheric parameters of field L and T dwarfs. *Astrophys. J.* **678**, 1372-1395 (2008).

**Acknowledgments: We** thank A. Conrad, S. Dahm, J. Lyke, H. Tran, H. Hershley, J. McIlroy, J. Rivera, and the entire Keck staff for maximizing our observing efficiency. We also thank J. Larkin and S. Wright for their assistance with OSIRIS data reduction, and R. Murray-Clay and K. Öberg for assistance with abundance values. Portions of this work were performed under the auspices of the U.S. Department of Energy by Lawrence Livermore National Laboratory under Contract DE-AC52-07NA27344. Support for this work was provided by NASA Origins of the Solar System grants to LLNL and Lowell Observatory and from the Keck Principal Investigator Data Analysis Fund, managed by NExScI on behalf of NASA. Support was also provided by the NASA High-End Computing (HEC) Program through the NASA Advanced Supercomputing (NAS) Division at Ames

Research Center. QMK is a Dunlap Fellow at the Dunlap Institute for Astronomy & Astrophysics, University of Toronto. The Dunlap Institute is funded through an endowment established by the David Dunlap family and the University of Toronto. The W.M. Keck Observatory is operated as a scientific partnership among the California Institute of Technology, the University of California and the National Aeronautics and Space Administration. The Observatory was made possible by the generous financial support of the W.M. Keck Foundation. The authors also wish to recognize and acknowledge the very significant cultural role and reverence that the summit of Mauna Kea has always had within the indigenous Hawaiian community. We are most fortunate to have the opportunity to conduct observations from this mountain.

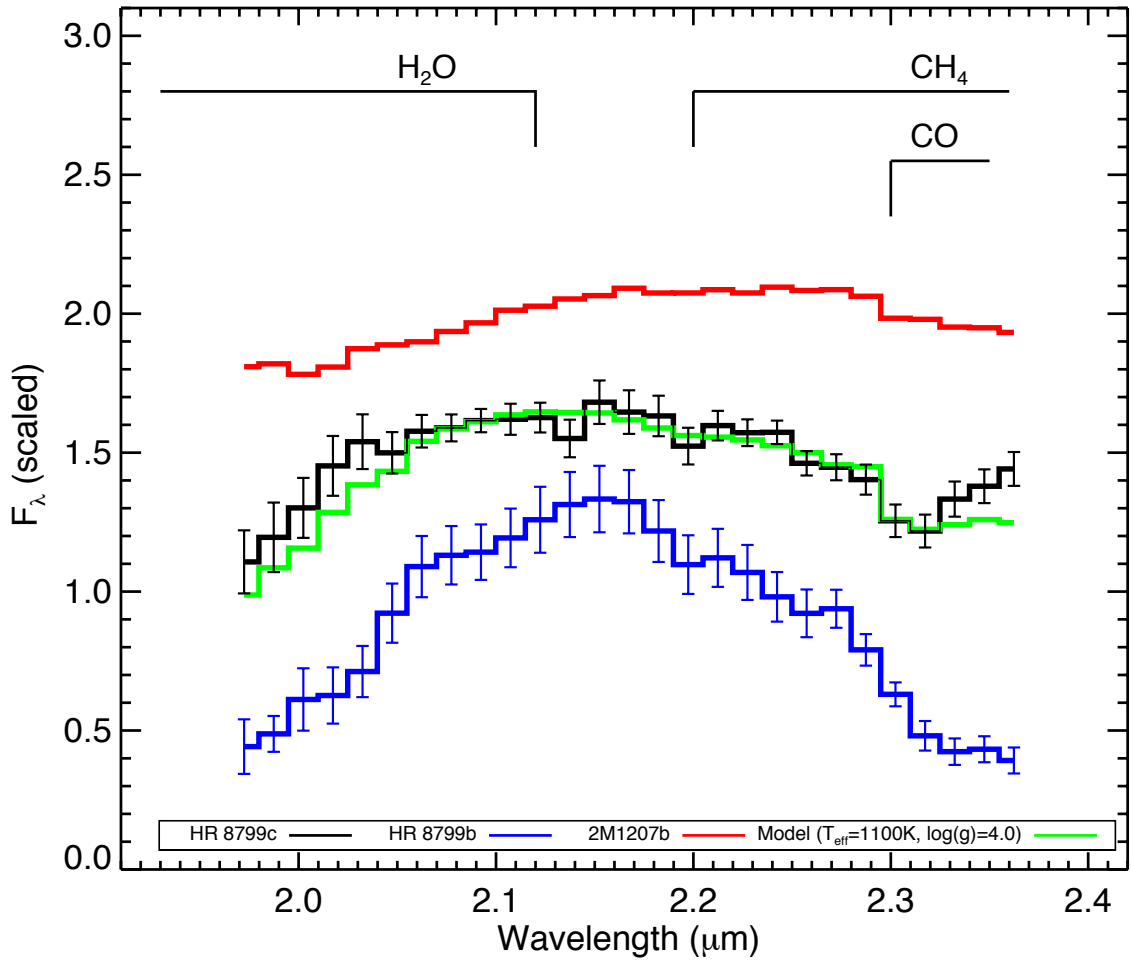

**Fig. 1**. Low resolution (λ/Δλ ~50) binned K-band spectrum for HR 8799c compared to the spectra of HR 8799b and 2M1207b. Binning the data bypasses issues related to the different resolutions and sampling among the three datasets while allowing us to compare broad molecular absorption bands (note that the residual speckle noise is correlated between the individual spectral channels, though this does not affect the analysis here). Each spectrum shows expected broad $H_2O$ absorption at λ < 2.15 μm that appears strongest in HR 8799b. While $CH_4$ and CO absorption likely contribute to the downward slope redward of ~2.15 μm for HR 8799b (*14, 40*),

their absorption is substantially weaker than is typical for an object near 1100 K. The spectrum of 2M1207b, however, has no discernible contribution from $CH_4$ and is likely shaped primarily by clouds, $H_2O$ and CO opacity (*7*). Shown in green is the best-fitting model spectrum for HR 8799c, with a temperature of 1100 K and a surface gravity of 100 m s$^{-2}$ (log (g) = 4.0, cgs units). The best fitting temperatures and gravities for HR 8799b and 2M1207b are ~750-1100 K and log(g) ~ 3.5-4.5 (*17*) and ~1000 K and log(g) ~ 4.0 (*7*), respectively.

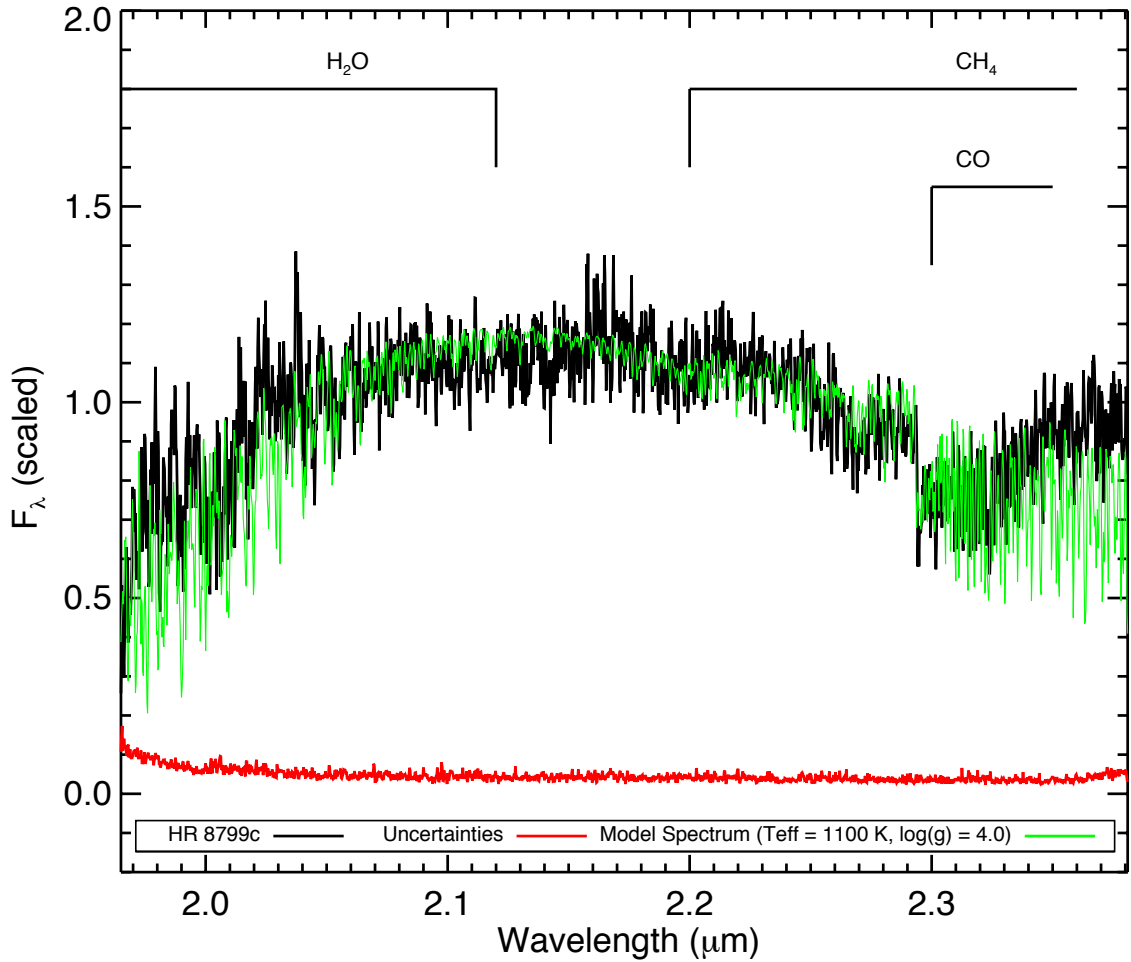

**Fig. 2.** K band spectrum of HR 8799c (black) at the full resolution of OSIRIS ($\lambda/\Delta\lambda$ ~4000). Both lines and continuum are shown, and the spectral features most relevant to objects of this mass are highlighted. For clarity, uncertainties are shown separately in red. Overplotted in green is the best fitting model spectrum. A clear drop from CO is detected, along with features typical of the CO bandhead at $\lambda > 2.29$ μm. The slight increase in the spectrum at the red end is attributed to residual speckle effects after the attenuation algorithm. Again, broad features from methane are not easily detected. Spectral data is provided in Database S1 (*15*).

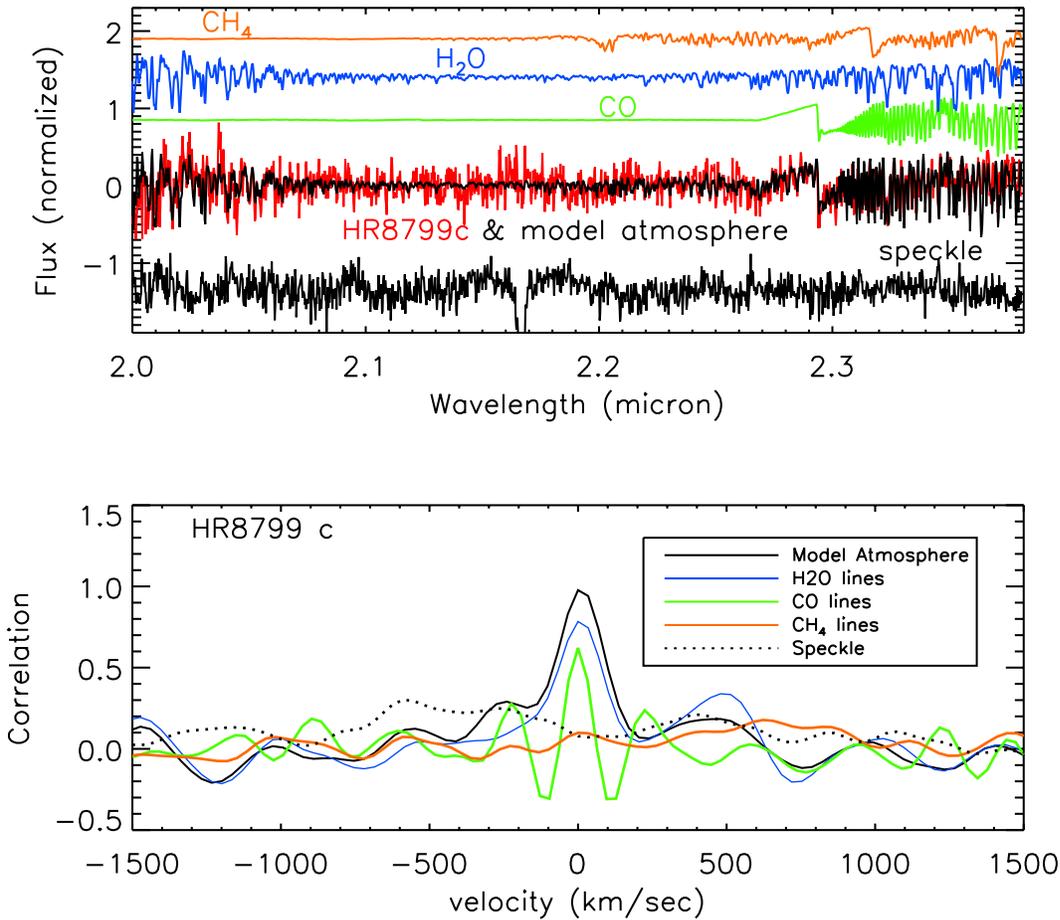

**Fig. 3. Top**: Pure $H_2O$ (blue), CO (green), and $CH_4$ (orange) synthetic spectra demonstrate the predicted location of absorption lines (*14*). The filtered spectrum of HR 8799c is shown in red. A filtered model atmosphere spectrum of mostly $H_2O$ and CO is overplotted in black. Also shown is a spectrum of a bright speckle, scaled such that the variance is equal to the variance in a featureless region of HR 8799c. **Bottom:** Cross-correlations functions for the spectrum of HR 8799c and the synthetic spectra shown in the top panel (solid) along with a baseline cross-correlation between the planet spectrum and a bright speckle (dotted). The cross-correlation with the $H_2O$-only template covered the entire observed wavelength range. Correlations with the CO and $CH_4$ templates were performed only over wavelength regions with strong lines ($CH_4$: $\lambda >$ 2.18 µm; CO: $\lambda >$ 2.29 µm). Significant cross-correlation peaks are found for the pure CO and $H_2O$ templates, as expected. The CO-only template produces two smaller symmetric peaks at ± 207 km s$^{-1}$ – this velocity corresponds to the near-equal line spacing of the CO lines across the (2,0) bandhead starting at 2.29 µm. Similar near-equal line spacing occurs for other CO bandheads (e.g. the (3,1) bandhead at 2.32 µm) resulting in the ringing behavior seen in the cross-correlation. No peaks in the correlations were found with either of the three $CH_4$ templates or the speckle spectrum (this is true for any subset of the observed wavelength range used). Cross-correlation is not required to detect molecular lines in our spectrum; however, this exercise

aids in quantifying the relative detections (or non-detection in the case of $CH_4$) when lines from individual molecules overlap.

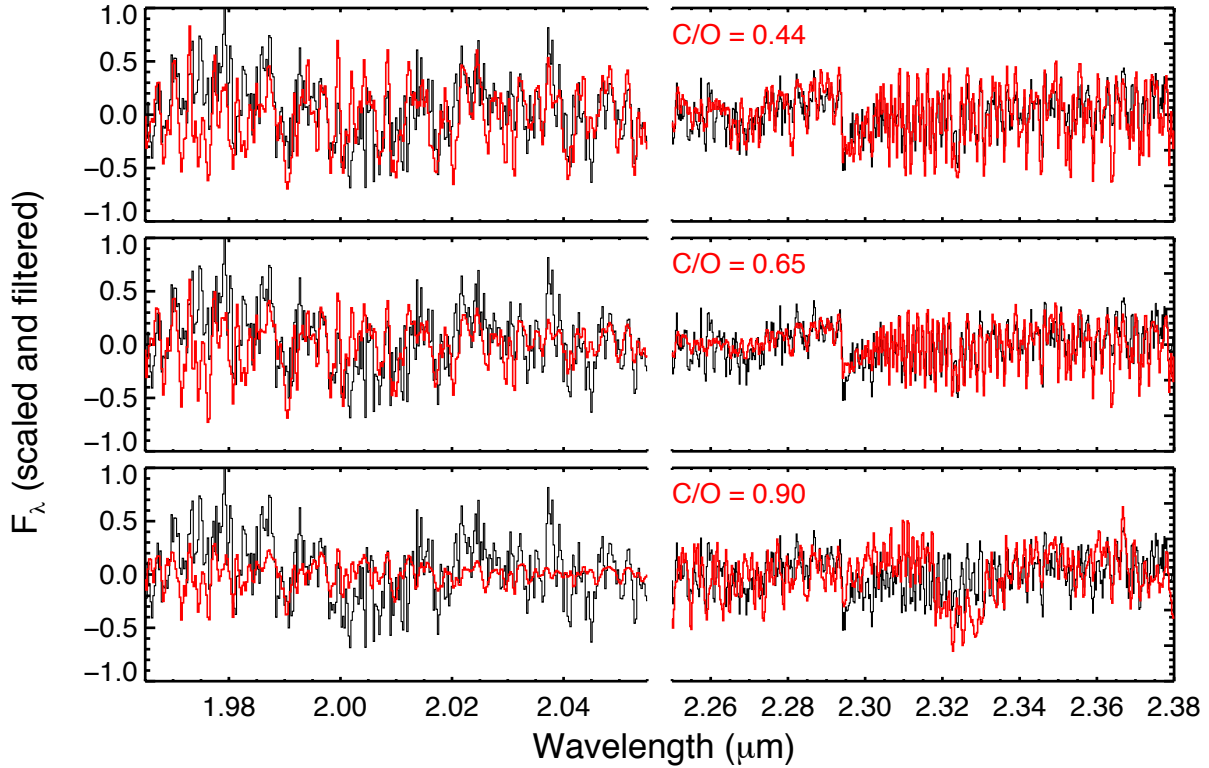

**Fig. 4.** Comparison of synthetic spectra (red) with different carbon and oxygen abundances to the low-pass filtered spectrum of HR8799c (black). The central wavelength range is omitted due to the lack of strong absorption features. Under the assumption that HR 8799c formed near its current location (i.e., little to no migration), the best matching carbon and oxygen abundances from the model grid are 8.33 and 8.51 (~0.06 and 0.13 below the solar values, using the traditional astronomical log base-10 abundance scale where the hydrogen abundance is 12.0), corresponding to C/O ~ 0.65 (middle panel). If HR 8799c somehow formed beyond 100 AU and migrated to its current location, the best matching C and O abundances are both substellar: 8.29 and 8.45, respectively, with C/O ~ 0.7 (not depicted). In either case, large C/O (~0.9, bottom panel) results in $CH_4$ increasing significantly and $H_2O$ decreasing in the spectrum, while for small C/O (~0.44, top panel), CO and $H_2O$ increase. Thus, both cases increase or decrease the prominent molecular lines identified in the K-band spectrum by factors that are easily excluded by the data (for instance, the large $CH_4$ absorption feature at 2.32 μm for C/O ~0.9). Note that the exclusion of the high and low C/O values is independent of $K_{zz}$ because the adopted value ($10^8$) always quenches the CO and $CH_4$ mole fractions at their maximum and minimum values, respectively. Raising $K_{zz}$ would have no impact and lowering $K_{zz}$ would only increase $CH_4$. All abundance values for C and O in the model grid are given in Tables S1 and S2.

**Supplementary Materials:**

Materials and Methods

Figures S1-S4

Tables S1-S2

Database S1

**Any Additional Author notes:** All authors contributed equally to this work.

# Supplementary Materials for

## Detection of Carbon Monoxide and Water Absorption Lines in an Exoplanet Atmosphere

Quinn M. Konopacky, Travis S. Barman, Bruce A. Macintosh, Christian Marois

correspondence to: konopacky@di.utoronto.ca

## Materials and Methods

### Spectral Extraction

The extraction of the spectrum of HR 8799c presents many challenges due to the brightness of the scattered starlight (speckles) at its location. The speckles are about a factor of 4 times brighter at 1" separation than they are at 1.7", the location of HR 8799b, making identification of the planet in each extracted data cube more difficult than in (*14*). Figure S1 shows example median-collapsed data cubes of HR 8799b from (*14*) and HR 8799c from this work. The two images are on the same scale. To first order, the position of these speckles varies linearly with wavelength (*41-42*) allowing them to be partially distinguished from the planet; however, chromatic effects in the imaging train break this simple behavior and make removal of the speckles complex.

We first use a spectral version of the locally optimized combination of images (LOCI, *43*) algorithm to identify the location of the planet in each cube. As in (*14*), this technique exploits the wavelength-dependent positional behavior of the speckles to facilitate their removal while leaving the non-spatially varying planet signal intact. We use a custom IDL routine that first smooths and rebins the data to $\lambda/\Delta\lambda$ ~50, then magnifies each slice of the cube at wavelength $\lambda$ about the position of the star by $\lambda_m / \lambda$, where $\lambda_m$ is 2.173 µm (the median wavelength in K broadband). This generates a cube in which the speckles are positionally aligned, but the planet position varies. Our spectral LOCI (SLOCI) takes as potential PSF images all slices of the cube that have sufficient separation in "wavelength" such that a signal from a real object does not self-subtract. This is somewhat analogous to the typical application of LOCI in angular differential imaging (ADI), where sufficient field rotation is required for frames to be used as a PSF (*43*). Because the field of view of the OSIRIS cubes is so small (0.32" x 1.28"), we use the whole image as the optimization region (7 x 28 $\lambda/D$). We found that when we did not mask the image while using SLOCI to subtract the speckle noise, the highest SNR identification of the planet's location was achieved. Once we have applied the algorithm and subtracted the speckles, we demagnify the cube, placing the planet signal back in its original location. This significantly reduces the speckle noise but also injects self-subtraction artifacts, much like the dark stripes seen in basic LOCI processing on either side of a real object. This unequivocally provides the location of the planet, but the self-subtraction makes extraction of an unbiased spectrum with this simple processing unreliable.

After using SLOCI to locate the planet in each data cube, the planet spectrum was extracted with the same algorithm used for HR8799b (*14*). Briefly, this method uses the same magnified cubes as described above, but then fits first order polynomials to the speckles as a function of wavelength. In this case, since we know the location of the planet, we mask it out so that it does not bias the polynomial fit. The results of these fits are then subtracted from the binned and full-resolution cubes before demagnifying. A collapsed version of a processed cube is shown in the right panel of Figure S1. The planet signal is extracted from both the binned and the full resolution cubes using a square 3x3 spatial pixel aperture centered on the location found using SLOCI. To obtain our final combined spectrum, we normalize each individual spectrum to the same mean flux to account for seeing and background variations, then median-combine all 33 individual spectra together. In the case of the full resolution spectra, we apply a barycentric correction to the individual spectra before combining them.

Uncertainties are calculated by finding the variance between the individual spectra at each wavelength. This uncertainty therefore includes contributions from both the statistical error in the flux of the planet and from the speckles. There is also some additional error in the blue end of the spectrum due to the large telluric features in this region (OH sky lines are well-subtracted and contribute negligibly to our uncertainties). The spectrum of the star itself is almost featureless except for the hydrogen Br $\gamma$ line. The stellar speckles therefore create broad variation with wavelength that primarily impacts the continuum morphology. Fits that rely on the continuum shape, for instance those that determine the best-fit effective temperature, are dominated by these correlated errors. The depth of narrow spectral features seen in the full-resolution spectrum, however, should be impacted primarily by Poisson noise (in the planetary and scattered star-light spectrum) as they are narrower than the speckle features. We therefore estimate the relative contributions of the speckle and Poisson uncertainties by generating smoothed versions of the full-resolution spectra that encapsulate the shape imparted by the speckles but not the narrow features from water and CO (window width of 200 at this resolution). The smoothed version of the spectrum is then subtracted from the data, leaving behind only the absorption features. This is somewhat analogous to applying a high-pass filter to the spectra, so we refer to these versions of the spectra as "filtered". As with the binned spectrum, we calculate the standard deviation of the smoothed and filtered spectra at each wavelength. The average uncertainty across all spectral channels in the raw full-resolution spectrum is about 8% of the measured fluxes. The correlated (with wavelength) error in the smoothed spectrum is on average about 6% while the average uncorrelated error (photon noise) is about 5%. Only the latter impacts out ability to fit the depth of individual lines.

The validity of our extraction method was extensively tested using a series of Monte Carlo simulations in which planets with flat spectra were injected into our data cubes at locations similar to that of the planet, and then re-extracted using our algorithms. This allowed us to confirm our uncertainty estimates and verify that we do not have any significant biases imparted from the speckle removal process. These tests show a slight undersubtraction of the speckle flux in the last few wavelength channels of the cubes. This is a minor effect (less than 2 times the noise estimated for the spectrum), but to avoid a bias, the wavelengths redward of 2.35 μm were excluded from the fits involving the continuum.

Derivation of Surface Gravity, Temperature, and C/O Ratio

The OSIRIS spectrum presented here provides constraints from both the continuum morphology at $\lambda/\Delta\lambda\sim50$, and the line depth and shape from the filtered spectrum, as described above. There is some degeneracy between gravity, temperature, and cloud thickness in the lines and continuum, with each data type (full spectral resolution, binned, and filtered) varying in sensitivity to each of these parameters. For example, at high-resolution the individual lines become deeper as the cloud height and location is decreased, while the line ratios do not change significantly. However, the K-band spectrum of a planet depends on the cloud properties, providing tight constraints on the viable model cloud parameters. We therefore fit the wavelength ranges and post-processed forms of our data that were the most sensitive to each of these atmospheric properties to provide an overall best fitting model that is consistent with all three data sets. We use broad band photometry from previous studies, spanning ~1 to 5 µm, (*2, 17-21*) to confirm that our best fits to the spectrum are consistent with the overall SED. The cloud properties are also well constrained by these broad band observations and our results agree with previous work (*21*).

As discussed in the main text, interior structure model constraints (1.0 $R_{Jup}$ < R < 1.5 $R_{Jup}$), dynamical stability arguments (M < 10 $M_{Jup}$), and system age (~30 Myr) provide a broad first cut in viable log(g): ~3.5 – 4.4. These constraints also limit the range of acceptable effective temperatures to 900 – 1300 K. We restricted our grid of atmosphere models to these ranges, with step sizes of 100 K in temperature and 0.5 in log(g), and use a simple parameterized cloud model (*14*). Our previous study of HR8799b (*14*) focused on solar abundance models but did explore a specific grid with $T_{eff}$ and log(*g*) consistent with hot-start evolution models and with all metals scaled to three and ten times the solar value. Uniformly enhanced metals (by factors of 3 and 10 over solar) could improve the model-observation comparison and consistency with cooling track predictions. However, after this study, the photometry for HR8799b changed significantly with the addition of an M-band (~4.67 µm) flux measurement (*21*) and a major revision of the observed flux at 3.3 µm (*18*). It is unlikely that the uniformly metal rich model will agree with the revised 3.3 µm data and has already been shown to be inconsistent with the M-band data (*21*). Therefore we have not considered the uniformly enhanced metallicity model here.

By fitting the binned and full resolution filtered spectra separately to the model grid using a similar method to (*44*) (with the weighting of all points set to one), we can determine which resolution can best determine temperature and gravity. We find that the depth of lines in the filtered spectrum has more distinguishing power in log(g) and the binned spectrum (similar to the SED) has more sensitivity to temperature. In particular, the depth of the first drop of the CO (2,0) bandhead is most sensitive to gravity for a given temperature (Fig. S2). We therefore elected to estimate gravity by comparing in detail the CO feature at ~2.294 µm to the model grid through two different methods.

First we considered only the ratio of the flux on either side of the band head. We compared the ratio in our measured spectrum with the ratio from the models in the grid (Fig. S3 top left) and find that the spectrum favors lower gravity models. We also performed a standard $\chi^2$ fit to the spectrum, using only a very narrow range encompassing the CO (2,0) bandhead drop (2.29 – 2.30 µm). The results of this fitting are shown in Figure S3 (top right). Again, the highest values of log(g) in the grid are disfavored. Using these two diagnostics, we conclude that the log(g) of HR 8799c is between 3.5 – 4.0.

In order to determine our best-fit temperature, we restricted our grid to only those surface gravities favored by the high resolution filtered spectrum. We again used a $\chi^2$ method to determine the preferred temperatures in the binned spectrum (K-band continuum). The resulting distribution of $\Delta\chi^2$ is shown in Figure S3 (bottom left). In this case, we take as the allowed range all parameters that have $\Delta\chi^2$ of 5 or less from the best-fit value. The best fitting temperature is 1100 K at log(g) = 4.0, but temperatures of 1000 K and 1200 K are also allowed. We therefore assign a temperature of 1100 ± 100 K to HR 8799c based on all available data. The data also favor a moderate cloud thickness, slightly thinner than that preferred for HR 8799b, consistent with the differences between their near-infrared colors. As mentioned in the main text, both the preferred temperature and gravity ranges are consistent with previous results (*2, 17-21*). Even when these parameters are unrestricted, the radius inferred from the best-fitting atmosphere model and the observed bolometric luminosity is consistent with hot-start evolution models (*17*).

We used these $T_{eff}$ and $\log(g)$ values for our exploration of C/O ratio. Two new model grids were computed using effective temperatures between 1000 – 1200 K and $\log(g)$ between 3.5 – 4.0. Cloud thickness and Kzz parameters were held fixed at their best-fit values. The carbon and oxygen abundances are given in Table S1 (consistent with HR8799c forming near its current location at 40 AU), and in Table S2 (corresponding to HR8799c having formed beyond 100 AU and migrated to its current location). These abundance values were chosen based on (*35*), which uses disk chemistry and a model for the process of core accretion to predict C and O abundances at locations of interest for the HR 8799 planets. For all model atmospheres, the equilibrium chemistry was recomputed (at every iteration) to insure that the global atmospheric properties of the final converged model (e.g. cloud composition, temperature-pressure profile, and opacities) are consistent with the elemental abundances. See Figure 3 of (*35*) for a graphical representation of the C abundance and C/O (relative to stellar) as a function of the solid accretion. For these calculations, we assumed that the C and O abundance of HR 8799A is solar, as found by (*34*). We note that if the C and O of the primary were non-solar, a new grid of planet C and O abundances based on (*35*) would need to be computed using these revised values. We therefore caution that our results below are based on this assumption.

As described in the main text, we performed $\chi^2$ fitting of the high resolution filtered spectrum to the grid. The best fitting C/O ratio (fixing the temperature and log(g) to 1100 K and 4.0) is 0.65 for formation at ~40 AU and 0.70 for formation beyond 100 AU. To determine the uncertainties in these values, we performed a Monte Carlo simulation where we resampled the spectrum using the uncorrelated uncertainties and fit the resampled spectra using the full range of plausible temperature and log(g). The spread in the allowed C/O values was dominated by the uncertainties in $T_{eff}$ and $\log(g)$ and found to be between 0.60 and 0.75, leading to a final best fit of $0.65^{+0.10}_{-0.05}$. Figure S4 shows the $\Delta\chi^2$ distribution resulting from this simulation. The returned C/O ratio is more sensitive to gravity than temperature. Higher gravity templates return lower C/O ratios at a given temperature. As mentioned above, the line depths are also impacted by cloud thickness. This parameter was held fixed in order to preserve the good match to the photometry and low-resolution spectrum. Nevertheless, potential uncertainty related to the cloud modeling is not captured in our Monte Carlo simulation.

It is important to note that different disk-chemistry models and formation scenarios could yield different C and O abundance pairs than explored here. For example, additional sources of carbonaceous material may have played a significant role during the formation of Jupiter (*37*) but are not explicitly considered in the (*35*) model. HR8799c orbits much farther from the $H_2O$ frost line than Jupiter and it is therefore unlikely that additional carbonaceous material would have a significant impact on C/O. If such material was important, the prediction would be high C/H and, thus, inconsistent with our data. Furthermore, the ability of atmosphere models to accurately retrieve individual elemental abundances has not been tested at very low temperatures. Previous model atmosphere comparisons to brown dwarfs and directly imaged planets have mostly assumed solar abundances and to our knowledge no one has measured the

C/O in field brown dwarfs. Also, very few cloudy objects are known that have effective temperatures as cool as the HR8799 planets. It is therefore difficult to assess how well the models retrieve C and O abundances for a larger population of objects and any systematic offset is not included in our Monte Carlo simulation and final error bars.

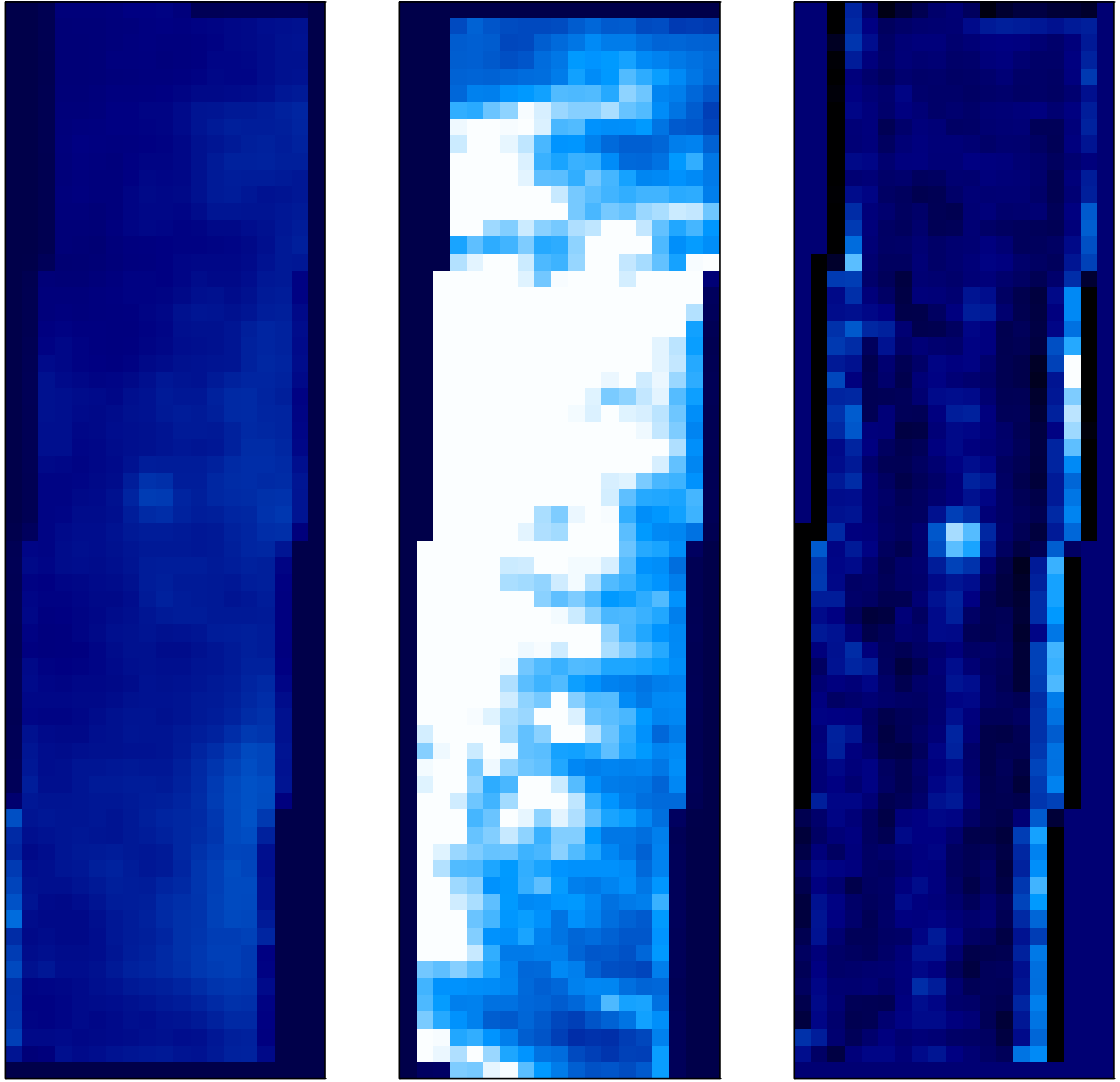

**Fig. S1.**

Collapsed, unprocessed OSIRIS spectral cubes of HR 8799b (**left**) and HR 8799c (**center**). The two images are on the same scale, and the planet location is near the center of each cube. While it is possible to see HR 8799b without any speckle removal, the brightness of the speckles at the separation of HR 8799c makes this extremely difficult (~4 times brighter than at the separation of HR 8799b). A collapsed cube to which our speckle suppression algorithm has been applied is also shown (**right**). The algorithm is able to substantially reduce the impact of the speckles, revealing HR 8799c near the center of the cube.

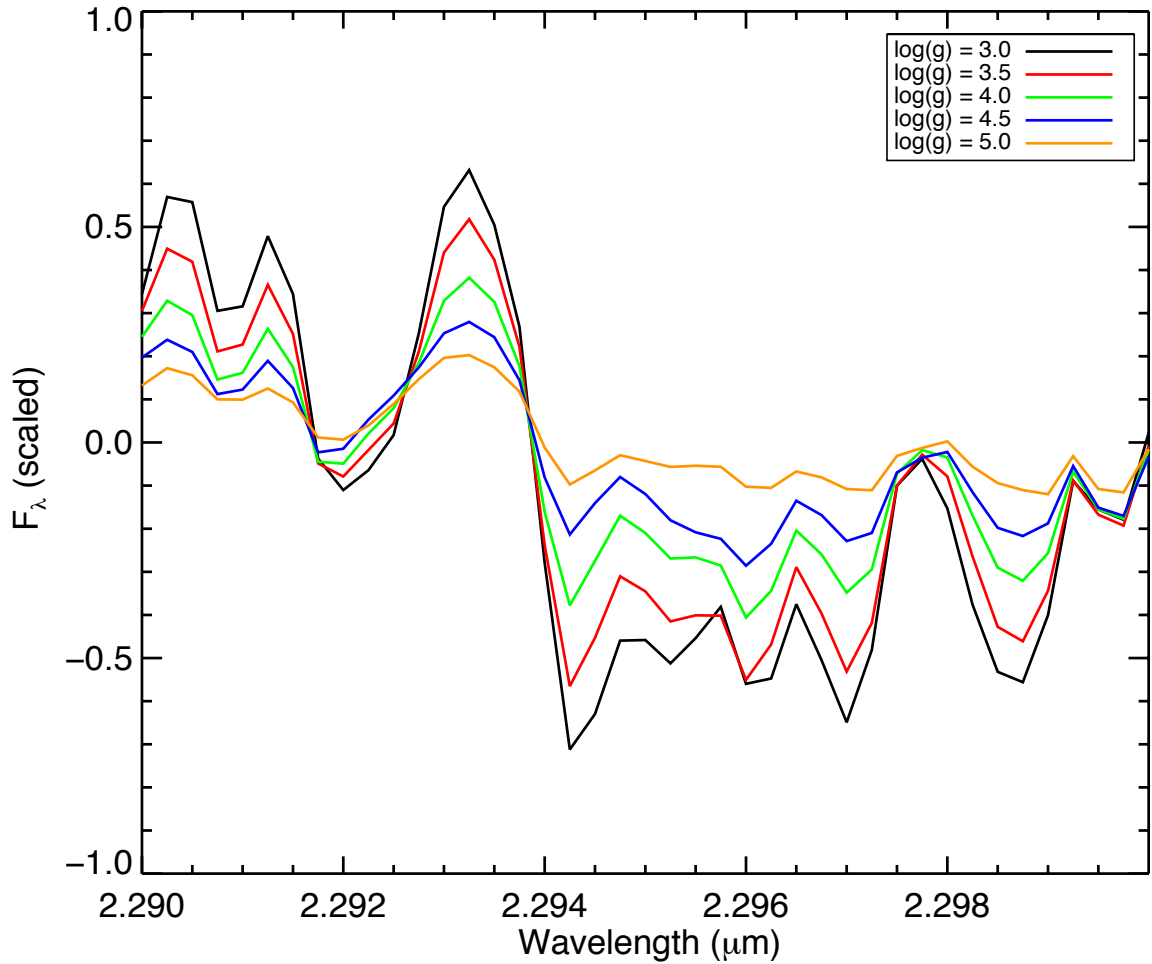

**Fig. S2**

PHOENIX model atmosphere spectra at T = 1100 K at the first drop of the CO (2,0) bandhead. Models from log(g) = 3.0-5.0 are shown to demonstrate the change in depth and morphology of these feature due to changing surface gravity. This spectral line has the most significant change due to surface gravity, much larger than the uncertainties on the spectrum of HR 8799c in this region. We therefore focused on this feature to determine the range of log(g) for HR 8799c.

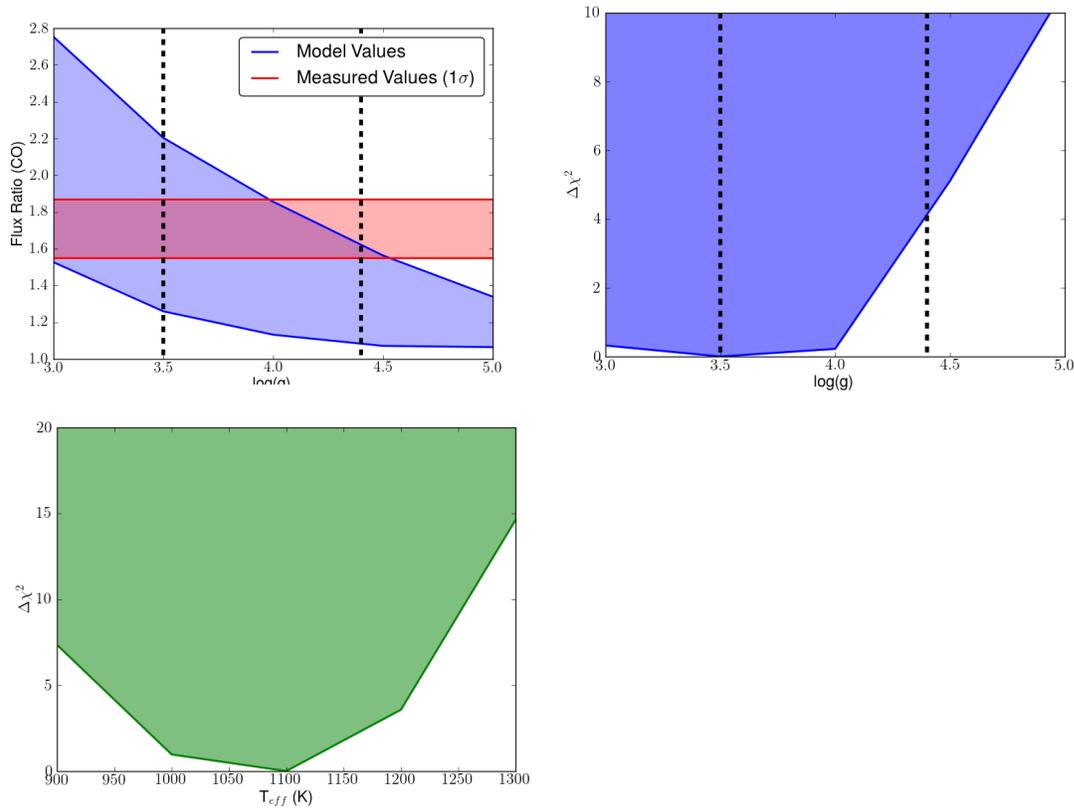

**Fig. S3**

**Top Left**: Distribution of flux ratios between the top and the bottom of the first drop in the CO (2,0) bandhead as measured in the full grid of spectral models (blue shaded region). The grid extends from log(g) = 3.0 – 5.0, so we have included the range from the full grid for completeness. At each gravity value, all grid temperatures (900-1300 K) and cloud parameters are included, generating the spread in flux ratio. However, in determining our final most plausible range of surface gravity, we limit the log(g) values allowed by other considerations (mass limits, age, luminosity, and radius). That range is bracketed in the figure by black dotted lines (3.5 – 4.4). Overplotted in red is the range of flux ratio values allowed within the 1σ uncertainties of the full resolution filtered spectrum of HR 8799c. There is overlap between the measurement and the model values across the full range of log(g) that we consider: however, the figure illustrates that high values of log(g) would be excluded without our limits. **Top Right:** Distribution of $\Delta\chi^2$ versus log(g) for fits across the first drop in the CO (2,0) bandhead (2.29 – 2.30 μm). We include $\Delta\chi^2$ values from the full model grid in the figure, but restrict ourselves to log(g) = 3.5 – 4.4 for determining the allowed range of values. The $\chi^2$ fitting prefers low gravity solutions, rising steeply after log(g) ~ 4.0. Given this and our lower cutoff of log(g) = 3.5, we refine the range of allowed surface gravity for HR 8799c to 3.5 – 4.0. **Bottom Left:** Distribution of $\Delta\chi^2$ versus temperature after restricting surface gravity to 3.5 – 4.0 based on fits to the binned (R ~ 50) spectrum. The best fit model has $T_{eff}$ = 1100 K, log(g) = 4.0. Given the large number of parameters needed to generate each model spectrum, we include in our range of allowed temperatures those that return a $\Delta\chi^2$ < 5. This gives us our final value of 1100 ± 100 K.

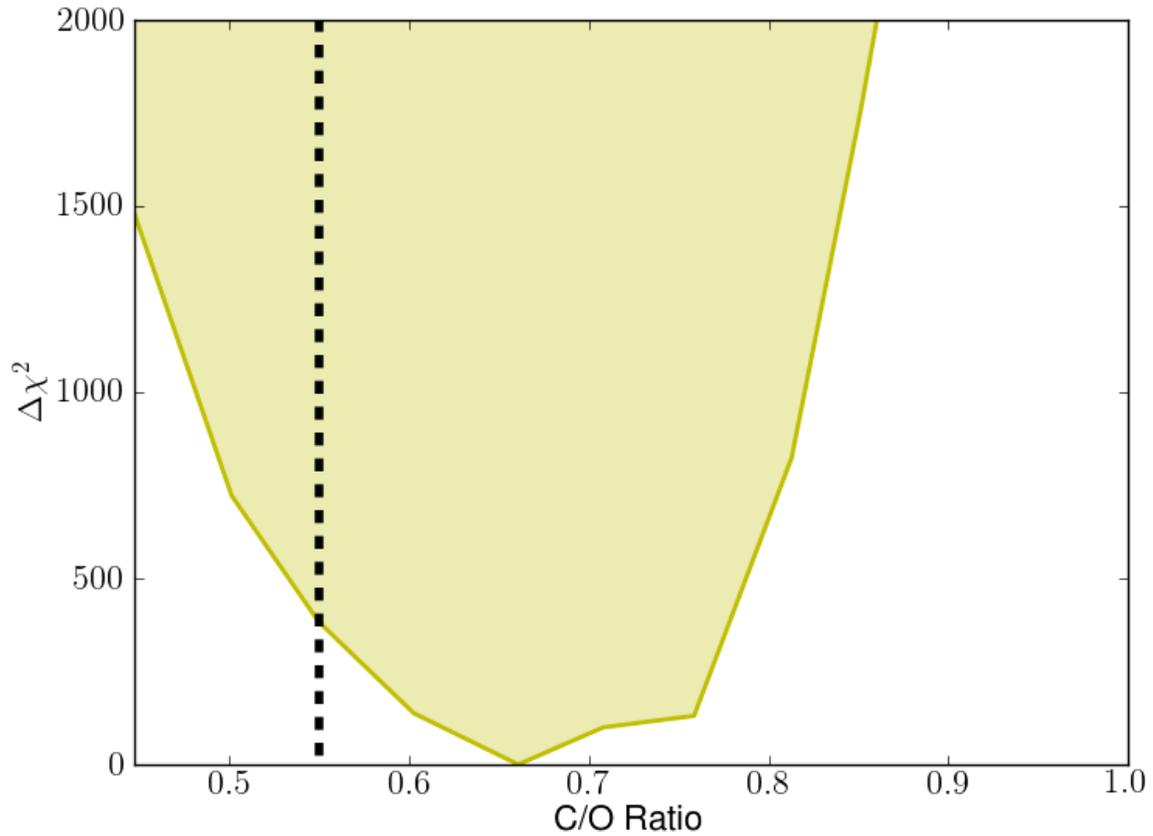

**Fig. S4**

Distribution of $\Delta\chi^2$ versus C/O ratio for the carbon and oxygen abundances given in Table S1 (appropriate if HR 8799c formed near its current location of ~40 AU), with surface gravity 3.5-4.0 and temperature $1100 \pm 100$ K. The best fit is C/O = 0.65. The dashed line denotes the stellar value. The distribution rises steeply below C/O ~ 0.6 and above C/O ~ 0.75. The spread in best-fit values, given the uncertainty in temperature and gravity, give C/O between 0.6 – 0.75. From these considerations, we arrive at our value of $0.65^{+0.10}_{-0.05}$.

**Table S1**.

Carbon and oxygen abundance values used to model atmospheres accreted between the $H_2O$ and $CO_2$ ice lines around HR 8799A (~10 – 100 AU). The values are derived from the prescription given in (*35*) assuming solar abundances of C and O in HR 8799A. This prescription defines the relative abundance of C and O by varying the percentage of solid versus gas accretion in the atmosphere from 0% (corresponding to the highest C/O) to 2% (the lowest C/O). HR 8799c currently resides at ~40 AU, implying that these are the most relevant values for C/O if it formed *in situ* with minimal migration. * ~ solar value.

| C | O | C/O |
|---|---|---|
| 8.48 | 8.82 | 0.45 |
| 8.42 | 8.72 | 0.50 |
| 8.38 | 8.64 | 0.55* |
| 8.35 | 8.57 | 0.60 |
| 8.33 | 8.51 | 0.65 |
| 8.31 | 8.46 | 0.70 |
| 8.29 | 8.42 | 0.75 |
| 8.28 | 8.37 | 0.80 |
| 8.26 | 8.33 | 0.85 |
| 8.25 | 8.30 | 0.90 |

**Table S2.**

Carbon and oxygen abundance values used to model atmospheres accreted between the $CO_2$ and CO ice lines around HR 8799A (>100 AU), again corresponding to solid accretion percentages ranging from 2% to 0%. These values are only relevant for HR 8799c in the case where it formed beyond 100 AU and migrated inward to its current location at ~40 AU. * ~ solar value.

| C | O | C/O |
|---|---|---|
| 8.49 | 8.84 | 0.45 |
| 8.43 | 8.73 | 0.50 |
| 8.38 | 8.64 | 0.55* |
| 8.34 | 8.56 | 0.60 |
| 8.32 | 8.50 | 0.65 |
| 8.29 | 8.45 | 0.70 |
| 8.27 | 8.40 | 0.75 |
| 8.26 | 8.35 | 0.80 |
| 8.24 | 8.31 | 0.85 |
| 8.23 | 8.28 | 0.90 |
| 8.22 | 8.24 | 0.95 |
| 8.22 | 8.21 | 1.00 |

**Database S1.**

ASCII file containing final combined spectrum at full OSIRIS resolution. Wavelength is given in microns. Flux and uncertainties are normalized, and are proportional to $F_\lambda$.